\documentstyle[12pt]{article}
\newskip\humongous \humongous=0pt plus 1000pt minus 1000pt

\newif\ifdtup

\def\ie{\hbox{\it i.e.}{}} 
 
\def\etal{\hbox{\it et al.}} 

\def\Tr{\mathop{\rm Tr}}

\def\abs#1{\left| #1\right|}
\def\pr#1{#1^\prime}

\def\beq{\begin{equation}}
\def\eeq{\end{equation}}

\def\beqn{\begin{eqnarray}}
\def\eeqn{\end{eqnarray}}
\relax

\def\dotx{\dotx{\dot\overline{x}}}

\relax

\catcode`\@=11

\@addtoreset{equation}{section}
\def\theequation{\thesection\arabic{equation}}

\def\@normalsize{\@setsize\normalsize{15pt}\xiipt\@xiipt
\abovedisplayskip 14pt plus3pt minus3pt%
\belowdisplayskip \abovedisplayskip
\abovedisplayshortskip \z@ plus3pt%
\belowdisplayshortskip 7pt plus3.5pt minus0pt}

\def\small{\@setsize\small{13.6pt}\xipt\@xipt
\abovedisplayskip 13pt plus3pt minus3pt%
\belowdisplayskip \abovedisplayskip
\abovedisplayshortskip \z@ plus3pt%
\belowdisplayshortskip 7pt plus3.5pt minus0pt
\def\@listi{\parsep 4.5pt plus 2pt minus 1pt
     \itemsep \parsep
     \topsep 9pt plus 3pt minus 3pt}}

\@twosidetrue

\relax

\catcode`@=12

\catcode`\@=11

\def\section{\@startsection{section}{1}{\z@}{3.5ex plus 1ex minus
   .2ex}{2.3ex plus .2ex}{\large\bf}}

\def\thesection{\arabic{section}.}

\def\appendix{\setcounter{section}{0}
 \def\thesection{APPENDIX \Alph{section}:}
 \def\theequation{\Alph{section}.\arabic{equation}}}

\def\ps@headings{\def\@oddfoot{}\def\@evenfoot{}
\def\@oddhead{\hbox{}\hfill
 \makebox[.5\textwidth]{\raggedright\ignorespaces --\thepage{}--
 \hfill {}}}  
\def\@evenhead{\@oddhead}
\def\subsectionmark##1{\markboth{##1}{}}
}

\ps@headings

\catcode`\@=12

\def\figcap{\section*{Figure Captions\markboth
 {FIGURECAPTIONS}{FIGURECAPTIONS}}\list
 {Fig. \arabic{enumi}:\hfill}{\settowidth\labelwidth{Fig. 999:}
 \leftmargin\labelwidth
 \advance\leftmargin\labelsep\usecounter{enumi}}}
 \relax
\def\tablecap{\section*{Table Captions\markboth
 {TABLECAPTIONS}{TABLECAPTIONS}}\list
 {Table \arabic{enumi}:\hfill}{\settowidth\labelwidth{Table 999:}
 \leftmargin\labelwidth
 \advance\leftmargin\labelsep\usecounter{enumi}}}
 \relax
\def\reflist{\section*{References\markboth
 {REFLIST}{REFLIST}}\list
 {[\arabic{enumi}]\hfill}{\settowidth\labelwidth{[999]}
 \leftmargin\labelwidth
 \advance\leftmargin\labelsep\usecounter{enumi}}}
 \relax

\catcode`\@=11

\def\ps@headings{\def\@oddfoot{}\def\@evenfoot{}
\def\@oddhead{\hbox{}\hfill
 \makebox[.5\textwidth]{\raggedright\ignorespaces --\thepage{}--
 \hfill {}}}    
\def\@evenhead{\@oddhead}
\def\subsectionmark##1{\markboth{##1}{}}
}

\ps@headings

\relax

\relax
\def\pl#1#2#3{{\it Phys. Lett. }{\bf #1}(19#2)#3}

\def\prep#1#2#3{{\it Phys. Rep. }{\bf #1}(19#2)#3}
\def\pr#1#2#3{{\it Phys. Rev. }{\bf #1}(19#2)#3}
\def\np#1#2#3{{\it Nucl. Phys. }{\bf #1}(19#2)#3}

\relax

\begin{document}
\newcommand\as{\alpha_S}
\newcommand\nf{{n_{\rm f}}}
\newcommand\refq[1]{$^{[#1]}$}
\newcommand\avr[1]{\left\langle #1 \right\rangle}
\newcommand\lambdamsb{
\Lambda_5^{\rm \scriptscriptstyle \overline{MS}}
}
\newcommand\ep{\epsilon}
\newcommand\half{{\scriptstyle \frac{1}{2}}}
\newcommand\epem{e^+e^-}
\newcommand\litwo{{\rm Li}_2}
\newcommand\qqb{{q\overline{q}}}
\newcommand\asb{\as^{(b)}}
\newcommand\qb{\overline{q}}
\newcommand\sigqq{\sigma_{q\overline{q}}}
\newcommand\siggg{\sigma_{gg}}
\newcommand\mqq{{\cal M}_{q\qb} }
\newcommand\mgg{{\cal M}_{gg} }
\newcommand\mqg{{\cal M}_{qg} }
\newcommand\fqq{f_{q\qb}}
\newcommand\fqqs{f_{q\qb}^{(s)}}
\newcommand\fqqp{f_{q\qb}^{(c+)}}
\newcommand\fqqm{f_{q\qb}^{(c-)}}
\newcommand\fqqpm{f_{q\qb}^{(c\pm)}}
\newcommand\fgg{f_{gg}}
\newcommand\fggs{f_{gg}^{(s)}}
\newcommand\fggp{f_{gg}^{(c+)}}
\newcommand\fggm{f_{gg}^{(c-)}}
\newcommand\fggpm{f_{gg}^{(c\pm)}}
\newcommand\fggtm{{\tilde f}_{gg}^{(c\pm)}}
\newcommand\fqg{f_{qg}}
\newcommand\epb{\overline{\epsilon}}
\newcommand\thu{\theta_1}
\newcommand\thd{\theta_2}
\newcommand\omxr{\left(\frac{1}{1-x}\right)_{\tilde\rho}}
\newcommand\omyo{\left(\frac{1}{1-y}\right)_{\omega}}
\newcommand\opyo{\left(\frac{1}{1+y}\right)_{\omega}}
\newcommand\lomxr{\left(\frac{\log(1-x)}{1-x}\right)_{\tilde\rho}}
\newcommand\MSB{{\overline{\rm MS}}}
\setcounter{topnumber}{10}
\setcounter{bottomnumber}{10}
\renewcommand\topfraction{1}
\renewcommand\textfraction{0}
\renewcommand\bottomfraction{1}
\begin{titlepage}
\nopagebreak
\vspace*{-1in}
\hfill hep-ph/9302211 \\
{\leftskip 11cm
\normalsize
\noindent
\newline
CERN-TH.6787/93 \newline

}

\vskip .6 cm
\begin{center}
{\Large \bf \sc
Small-Size Instanton Corrections }

{\Large \bf \sc to the $\tau$ Hadronic Width}
\vskip .6cm
{\large \bf Paolo Nason\footnotemark}
\footnotetext{On leave of absence from INFN, Sezione di Milano, Italy.}
\vskip .3cm
{CERN TH-Division, CH-1211 Geneva 23, Switzerland}\\
\vskip .6cm
{\large \bf Massimo Porrati\footnotemark}
\footnotetext{On leave of absence from INFN, Sezione di Pisa, Italy.}
\vskip .3cm
{Department of Physics, New York University, New York, NY 10003, USA}\\
\end{center}
\vskip .6cm
\nopagebreak
\begin{abstract}
{\small
We compute the effect of small-size instanton corrections to
current-current correlators, for all combinations of axial, vector,
and possibly flavour non-diagonal currents.
We apply our result to the hadronic decays of the $\tau$ lepton,
in order to assess the reliability of the
determination of $\as$ from
the $\tau$ hadronic width.
}
\end{abstract}
\vfill
CERN-TH.6787/93 \newline
NYU-TH.92/11/03 \newline
January 1993    \hfill
\end{titlepage}
\section{Introduction}
In ref.~[\ref{Braaten}], the possibility of determining
$\as$ to high precision from the hadronic decay width of the $\tau$ lepton
has been put forward.
It has been argued that, in spite of the small mass of the $\tau$, this
determination is
perturbative, since in the contest of the operator product
expansion and QCD sum rules\refq{\ref{Shifman}} non-perturbative corrections
to this process
can be estimated to be small. This is essentially a consequence
of the fact that in the chiral limit the dominant non-perturbative corrections
to this quantity behave like $1/m_\tau^4$.

This determination presents several subtle points.
First of all, it relies on assumptions that are somewhat
 stronger than the usual
assumptions of perturbative QCD. One should in fact assume that it is
legitimate
to add to a truncated perturbative expansion, which is essentially an
expansion in inverse powers of logarithms of the momentum scale of the process,
terms that are formally suppressed by powers of the momentum scale itself.
A justification of this procedure can be found in
ref.~[\ref{Mueller}], where its validity is related to the position of the
singularities of the Borel transform of the perturbative expansion. There it is
also made clear that the underlying assumption of this procedure is that the
only Borel singularities which are present are the known ones,
\ie\ infrared renormalons and instanton singularities.

Recently some authors\refq{\ref{Sakarov}} have brought arguments
in favour of corrections, due to ultraviolet renormalons,
suppressed by two powers of the momentum
in current-current correlators. These corrections would give rise to terms
behaving like $1/m_\tau^2$ in the $\tau$ hadronic width, thereby spoiling
the analysis of ref.~[\ref{Braaten}].

In the present work we will not deal with the general validity of the
operator product expansion, or of the QCD sum rules formalism.
We will instead deal with the well-known fact\refq{\ref{Shifman}} that
instantons do spoil the operator product expansion by introducing corrections
that are power-suppressed by 9 or more inverse powers of the momentum.
In view of the large power suppression, it is clear that corrections
of this type will behave more like a step function, below which the
perturbative methods will certainly be inapplicable. In other words,
we expect a value of the ratio $m_\tau/\Lambda_{\rm QCD}$ below which instanton
corrections are of order one, and above which they are essentially zero.
We point out that this is the only non-perturbative correction that
can be computed explicitly in terms of the QCD parameter $\Lambda$
and the quark masses alone,
with no need of further phenomenological inputs.
Computations of this type of corrections have been carried out in refs.
[\ref{Gross}] and~[\ref{Ellis}] in the context of $e^+e^-\to\mbox{hadrons}$.
Although intermediate formulae
and results are the same in the two papers, the conclusions are
different, since the authors of ref.~[\ref{Gross}] conclude that instanton
corrections are suppressed by four powers of the momentum, while
in ref.~[\ref{Ellis}] the effects are suppressed by twelve powers.
For our purposes we are interested in the kind of result given in
ref.~[\ref{Ellis}], since the corrections mentioned in ref.~[\ref{Gross}]
should be viewed rather as instanton corrections to the value of the
matrix elements of dimension four operators,
which is usually determined empirically, and is not calculable from
first principles.

In ref.~[\ref{Ellis}], it is claimed that instanton corrections in
$\epem\to\;$hadrons become large for annihilation energies between 1 and
2~GeV.
If a similar result held in the case of the $\tau$, it would spoil
the analysis of ref.~[\ref{Braaten}].

In order to obtain results that can be applied to our case we must complete
the calculations of refs.~[\ref{Gross}] and [\ref{Ellis}] in three respects.
First of all, we need to include the corrections to the axial current
correlators, which were not considered there. Secondly, we must consider
non-diagonal currents for flavours of different masses, which also were not
considered previously.
Lastly, we must perform an analysis in a definite subtraction scheme,
in order to be able to relate the value of $\Lambda$ that we intend to use
in this context with the value that is extracted from high-energy experiments.
In fact, instanton effects are suppressed by a factor of
order $\exp(-2\pi/\as)$, so that a scheme redefinition,
which changes the inverse of $\as$ by the addition of a constant, affects
directly the prefactor of the instanton correction (at the time when
refs.~[\ref{Gross}] and [\ref{Ellis}] were written, there was no sound
agreement on the allowed range of $\Lambda_{\rm QCD}$, so that a complete
answer correctly including the scheme dependence would not have been very
useful). Our result also differs quantitatively from ref.~[\ref{Ellis}],
and we will comment upon the differences in due time.

Our paper is organized as follows:
in section~2 we give a description of the calculation of the process
in question. In section~3 we review the formula for the instanton density
in QCD, and specify its form in the $\MSB$ scheme.
In section~4 we apply our results
to the $\tau$ hadronic width. In section~5 we give our conclusions.
\section{Instanton Corrections to the \newline Current-Current Two-Point
         Functions}
We consider the correlator
of two currents in the instanton background
\beq
\label{corr1}
\Pi_{\mu\nu}(x,y;\Omega_\pm)=\langle J^{ud}_\mu(x)
J^{du}_\nu(y)\rangle_{A(\Omega_\pm)}
=-\Tr\left(\Gamma_\mu S^d_\pm(x,y;\Omega_\pm)\Gamma_\nu
           S^u_\pm(y,x;\Omega_\pm)\right),
\eeq
where we have
\beq
J^{ud}_\mu = \bar\psi^u \Gamma_\mu \psi^d,\quad\quad
J^{du}_\nu = \bar\psi^d \Gamma_\nu \psi^u;
\eeq
$\Gamma_\mu$ stands for $\gamma_\mu$ for vector currents,
and for $\gamma_5\gamma_\mu$ for axial current;
$\Omega_\pm$ denotes the instanton
(anti-instanton) global coordinates $\Omega_\pm=(z,\rho,R)$, where $z$ is
the position, $\rho$ is the size, and $R$ stands for the colour orientation.
$S_{\pm}$ denotes the fermion propagator in the instanton
(anti-instanton) background, whereas $S_0$ is the free fermionic propagator.
The superscript $u,d$ specifies the mass of the propagator. When no superscript
is given, it is massless.

We consider here the general case of non-diagonal flavour currents.
The two flavours involved,
which will be simply called $u$ and $d$, have different masses $m_u$ and
$m_d$.
The propagators in the instanton background have the small mass
expansion
\beqn
S^{u(d)}_\pm(x,y;\Omega_\pm)&=&-\frac{\Psi_0(x)\Psi_0^\dagger(y)}{m^{u(d)}}
\,+\,S_\pm(x,y;\Omega_\pm) \nonumber \\
&&+\,
m^{u(d)}\int dz^4 S_\pm(x,z;\Omega_\pm)S_\pm(z,y;\Omega_\pm)+{\cal O}(m^2),
\label{propagator}
\eeqn
where terms with odd powers of the mass commute with $\gamma_5$, while
the terms with even powers anticommute.

 From eqs.~(\ref{corr1}) and (\ref{propagator}) we
can immediately obtain the correlator in the limit of small masses
\beqn \label{pimunuinst}
\label{pidefV}
\Pi^{V}_{\mu\nu}&=& -\Tr(\gamma_\mu S_0(x,y) \gamma_\nu S_0(y,x))
 + A_{\mu\nu} + C B_{\mu\nu}
\\
\label{pidefA}
\Pi^{A}_{\mu\nu}&=& -\Tr(\gamma_\mu S_0(x,y) \gamma_\nu S_0(y,x))
 + A_{\mu\nu} - C B_{\mu\nu},
\eeqn
where $V$ stands for vector-vector and $A$ stands for axial-axial
correlators, and
\beqn
C &=& \frac{1}{2} \left(\frac{m_u}{m_d}+\frac{m_d}{m_u}\right) \nonumber \\
A_{\mu\nu} &=& \Tr(\gamma_\mu S_0(x,y) \gamma_\nu S_0(y,x))
-\Tr\left[\gamma_\mu
S_\pm(x,y;\Omega_\pm)\gamma_\nu S_\pm(y,x;\Omega_\pm)\right] \\
B_{\mu\nu} &=& 2\Tr\left[
\gamma_\mu\Psi_0(x)\Psi_0^\dagger(y)\gamma_\nu\int d^4 z S_\pm(x,z;\Omega_\pm)
S_\pm(z,y;\Omega_\pm)\right].
\eeqn
Expressions for the coefficients $A_{\mu\nu}$ and $B_{\mu\nu}$ have been
obtained in ref.~[\ref{Gross}]. Their result agreed with that of
ref.~[\ref{Ellis}].
We do however find a different colour factor normalization for the result,
our normalization being 2/3 of theirs.
We get (after colour averaging)
\beqn
&& A_{\mu\nu}(x,y,\rho)=\frac{1}{2\pi^4}S_{\mu\alpha\nu\beta}
\Bigg[\frac{\rho^4(h_xh_y)^2}{\Delta^4}(2\Delta^\alpha\Delta^\beta-
g^{\alpha\beta}\Delta^2) \nonumber \\
&&
+\frac{\rho^2}{\Delta^4}h_xh_y
\left(h_y(\Delta^\alpha y^\beta+\Delta^\beta y^\alpha)-
      h_x(\Delta^\alpha x^\beta+\Delta^\beta x^\alpha)\right) \Bigg]
+\mbox{odd terms}
\eeqn
\beq
 B_{\mu\nu}(x,y,\rho)=-\frac{1}{\pi^4}(h_x h_y)^2\frac{\rho^2}{\Delta^2}
\left[(\rho^2+x\cdot y)g_{\mu\nu}+(y_\mu x_\nu-x_\mu y_\nu)\right]
+\mbox{odd terms},
\eeq
where
\beqn &&
 h_x=1/(x^2+\rho^2),\quad \quad h_y=1/(y^2+\rho^2) \nonumber \\ &&
\Delta=x-y, \nonumber \\ &&
S_{\mu\alpha\nu\beta}=g_{\mu\alpha}g_{\nu\beta}-g_{\mu\nu}g_{\alpha\beta}
+g_{\mu\beta}g_{\nu\alpha}.
\eeqn
``Odd terms'' refers to terms that change sign when going from
an instanton to an anti-instanton, and thereby vanish when summing over
the two contributions. The instanton location has been chosen in $x_\pm=0$
in the above formulae.

The expression for the current-current correlator
including instanton corrections in the dilute-gas approximation is
\beq
\Pi_{\mu\nu}(\Delta)=
\Pi^0_{\mu\nu}(\Delta)+\int d^4z d\rho D(\rho)
\sum_\pm \left(\Pi_{\mu\nu}(x,y,z,\rho,\pm)-\Pi^0_{\mu\nu}(\Delta)\right),
\eeq
where $D(\rho)$ is the instanton density, which will be specified in
the following section.
Using eqs.~(\ref{pidefV}) and (\ref{pidefA}) we obtain
\beq \label{Pimunuform}
\Pi_{\mu\nu}(\Delta)=
\Pi^0_{\mu\nu}(\Delta)+\int d^4 z d\rho D(\rho)
2\left(A_{\mu\nu}(x-z,y-z,\rho)+ C B_{\mu\nu}(x-z,y-z,\rho)\right)
\eeq
for vector currents, and the same form with $C\to -C$ for axial
currents.
We have
\beq
\Pi^0_{\mu\nu}(\Delta) =
\frac{12 S^{\mu\alpha\nu\beta}\Delta_\alpha \Delta_\beta}{(2\pi^2)^2\Delta^8}.
\eeq
This formula includes the factor 3 from the colour trace.
Defining
\beqn \label{amunudef}
a_{\mu\nu}(\Delta,\rho)&=&\int d^4 z A_{\mu\nu}(x-z,y-z,\rho)
\\ \label{bmunudef}
b_{\mu\nu}(\Delta,\rho)&=&\int d^4 z B_{\mu\nu}(x-z,y-z,\rho),
\eeqn
the $z$ integration gives
\beqn
a_{\mu\nu}(\Delta,\rho)&=& -\frac{1}{2\pi^2}
\left[\frac{\partial^2}{\partial \Delta^\mu \partial \Delta^\nu }
G(\Delta^2,\rho) +
 2 G'(\Delta^2,\rho)  g_{\mu\nu}\right]  \label{amunu}\\
b_{\mu\nu}(\Delta,\rho)&=& \frac{1}{2\pi^2}
\left[\frac{\partial^2}{\partial \Delta^2 } G(\Delta^2,\rho)  +
 2 G'(\Delta^2,\rho) \right]  g_{\mu\nu}, \label{bmunu}
\eeqn
where $G$ and $G'$ are functions of $\Delta^2$ and $\rho$, and they are
given by the expressions
\beq
G'(\Delta^2,\rho) = \frac{\partial G(\Delta^2,\rho)}{\partial \Delta^2} =
\frac{\rho^2}{\Delta^4}\left[-\frac{2\rho^2}{\Delta^2}\frac{1}{\xi}
\log\frac{\xi-1}{\xi+1} - 1 \right],
\eeq
with $\xi=\sqrt{1+4\rho^2/\Delta^2}$.
The above expression is a non-analytic function of $\rho$ for small
$\rho$, while it is analytic in $\rho^{-1}$ for $\rho\to \infty $:
\beqn
\lim_{\rho\to 0} G'(\Delta^2,\rho) &=& \
\frac{\rho^2}{\Delta^4}
\left[-1-\frac{2\rho^2}{\Delta^2}\log\frac{\rho^2}{\Delta^2} +
...\right]
\\ \label{roinflim}
\lim_{\rho\to \infty } G'(\Delta^2,\rho) &=& \
-\frac{1}{6\Delta^2}
+\frac{1}{30\rho^2}
-\frac{\Delta^2}{140\rho^4}
+\frac{\Delta^4}{630\rho^6}
-\frac{\Delta^6}{2772\rho^8} ...\quad.
\eeqn

We need the integral of $G$ against the instanton density, which is given by
\beq
D(\rho)= H \left[\log\frac{1}{\rho^2\Lambda^2}\right]^c
\rho^M,
\eeq
where $M=6+\nf/3$.
The value of $H$ and of the power of the logarithmic term $c$
will be discussed in more detail in the next section.
We notice that the small-$\rho$ region never poses a convergence problem.
On the other hand, the large-$\rho$ region is divergent.
However, because of the
analiticity of the integrand in that region, the divergence will be
limited to a finite number of terms in the expansion of $G$.
For definiteness, let us assume that $\nf=3$. Then $D(\rho)$ will
behave like $\rho^7$ for small $\rho$. Then we will have to subtract from
$a_{\mu\nu}$ and $b_{\mu\nu}$ all the terms of their expansion up to the
power $\rho^{-8}$. These subtracted terms will all have power-like
dependence upon $\Delta^2$, with undefined, infrared divergent coefficients.
We interpret these infrared divergent terms as the instanton contribution
to the expectation value of operators in the operator product expansion
of the two currents we are considering.
These matrix elements are in general uncalculable, and they are
usually estimated on the basis of some phenomenological considerations.
The first term of this kind is in fact related to
operators of dimension four.
Since we want instead the true, direct effect of
the instanton upon our amplitude, we will subtract at the end
these divergent terms.
We first define
\beq
\tilde{G}'(\Delta^2,\rho,\rho_0)=G'(\Delta^2,\rho) -
\theta(\rho-\rho_0)
\frac{1}{\Delta^2}\sum_{j=0}^L g_j \left(\frac{\Delta}{\rho}\right)^{2j}
\label{gtildapdef}
\eeq
\beq
\tilde{G}(\Delta^2,\rho,\rho_0)={G}(\Delta^2,\rho) -
\theta(\rho-\rho_0) \left[\log\Delta^2\;g_0+
\sum_{j=1}^L \frac{g_j}{j} \left(\frac{\Delta}{\rho}\right)^{2j} \right],
\label{gtildadef}
\eeq
where the $g_j$ are the numerical coefficients in the expansion
of eq.~(\ref{roinflim}). The functions $\tilde{G}$, $\tilde{G}'$ are
then integrable
in $\rho$ in the whole range, even when multiplied by a power of $\rho^M$
with $M<2L-1$. Once we know that their integral is in fact
convergent, we may regulate it in any way we like. For example, we
may choose the analytic continuation method of taking $-5<M<-3$, and then
continuing to all the allowed values of $M$. We now compute the Mellin
transforms. We get
\beqn
\int_0^\infty d\rho \rho^M \tilde{G}'(\Delta^2,\rho,\rho_0)
&=&-\Delta^{M-1}\Gamma(-M-4)
\Gamma^2\left(\frac{M+5}{2}\right)
\sin\left(\frac{M\pi}{2}\right) \nonumber \\
&-&\frac{1}{\Delta^2}\sum_{j=0}^{L}
    g_j \frac{\Delta^{2j}\rho_0^{M-2j+1}}{M-2j+1}.
\label{gtildapmt}
\\
\int_0^\infty d\rho \rho^M \tilde{G}(\Delta^2,\rho,\rho_0)
&=&-\frac{2\Delta^{M+1}}{M+1}\Gamma(-M-4)
\Gamma^2\left(\frac{M+5}{2}\right)
\sin\left(\frac{M\pi}{2}\right) \nonumber \\
&-&\log\Delta^2\; g_0\frac{\rho_0^{M+1}}{M+1}-\sum_{j=1}^{L}
    \frac{g_j}{j} \frac{\Delta^{2j}\rho_0^{M-2j+1}}{M-2j+1}.
\label{gtildamt}
\eeqn
Observe that the above formulae are now convergent for $M<2L-1$. The
poles in $M$ arising from the gamma functions cancel against those
in the sums.
 From eq.~(\ref{Pimunuform}), (\ref{amunudef}), (\ref{bmunudef}),
(\ref{amunu}) and (\ref{bmunu})
we get
\beqn
&& \Pi_{\mu\nu}(\Delta)=\Pi^0_{\mu\nu}(\Delta)-
 H\left[\log\frac{1}{\Delta^2\Lambda^2}\right]^c \Gamma(-M-4)
\Gamma^2\left(\frac{M+5}{2}\right)
\sin\left(\frac{M\pi}{2}\right)
\frac{1}{\pi^2} \phantom{aaaaa}
\nonumber \\
&&\left\{
\left[C\frac{\partial^2}{\partial\Delta^2} g_{\mu\nu}
-\frac{\partial^2}{\partial\Delta^\mu\partial\Delta^\nu}\right]
\frac{2\Delta^{M+1}}{M+1}
+(C-1)2 g_{\mu\nu}\Delta^{M-1}
\right\}
\nonumber \\
&&-\int_0^{\rho_0}D(\rho) d \rho \frac{g_0}{\pi^2} \left\{
\left[C\frac{\partial^2}{\partial\Delta^2} g_{\mu\nu}
-\frac{\partial^2}{\partial\Delta^\mu\partial\Delta^\nu}\right]
\log\Delta^2
+(C-1)2 g_{\mu\nu}\frac{1}{\Delta^2} + P_{\mu\nu}(\Delta^2)
\right\}, \nonumber \\
\label{pimunudelta}
\eeqn
where $P_{\mu\nu}(\Delta)$ is a polynomial in $\Delta$.
Observe that the logarithmic power in the instanton
density is simply replaced by its value for $\rho=p$. Corrections
to this replacement are logarithmically suppressed, and therefore
are not included here.

The Fourier transform of the expression (\ref{pimunudelta})
can also be performed
for non-integer $M$, and then continued to the desired value of $M$,
using the formula
\beq \label{fouriert}
\int e^{i\Delta\cdot p} \Delta^J d^4 \Delta=
p^{-J-4} \frac{4\pi}{\Gamma\left(-\frac{J}{2}\right)} \cos\frac{J\pi}{2}
\Gamma\left(\frac{3}{2}\right)\Gamma\left(-\frac{J+3}{2}\right)\Gamma(J+4).
\eeq
Observe that this formula vanishes when $J$ is an even integer.
In fact, in this case the result is a distribution concentrated at $p^2=0$.
The term $P_{\mu\nu}(\Delta)$ therefore does not contribute to the
Fourier transform, and we get the result
\beqn
\Pi_{\mu\nu}(p^2)&=&\Pi^0_{\mu\nu}(p^2)
+ H\left[\log\frac{p^2}{\Lambda^2}\right]^c
\frac{(M+3)\Gamma\left(\frac{3}{2}\right) \Gamma^3\left(\frac{M+3}{2}\right)}
{(M+1)\Gamma\left(\frac{M+6}{2}\right)}
\nonumber \\
&& p^{-M-3}\left\{
(M+3)\left[C g_{\mu\nu}-\frac{p_\mu p_\nu}{p^2} \right]
+(C-1) g_{\mu\nu} \right\}
\nonumber \\
&+&\int_0^{\rho_0} D(\rho) d\rho\;
\frac{4}{3 p^4}\left[(1-3C)(g_{\mu\nu}p^2-p_\mu p_\nu)+3(1-C)p_\mu
p_\nu \right].
\label{finalpi}
\eeqn
The last term is explicitly infrared divergent for $\rho_0 \to \infty$.
We interpret this term as the contribution of the instanton to the
matrix elements of the dimension-four operators $F^2$ and $m\bar{\psi}\psi$.
There are no singularities in the remaining terms
for positive values of $M$.

It is easy to check that, in the particular
case of flavour-diagonal vector currents, the $1/p^4$ term corresponds to
the term obtained in refs.~[\ref{Gross}] and [\ref{Ellis}].
One can argue that the $1/p^4$ term cannot contribute to the
discontinuity of the polarization operator, and therefore it does not
affect the cross sections for hadron production.
This formally correct argument fails however when radiative corrections are
included. Similarly, we have neglected a number of contributions
localized at $p^2=0$. Radiative corrections may delocalize these
contributions, and therefore give corrections to the discontinuity,
which behave like three or more inverse powers of $p^2$.
These terms cannot be neglected, and, being IR-divergent, cannot
be computed by perturbative techniques. They are usually accounted for
phenomenologically, when the value of the various
condensates is extracted from data.

\section{The Instanton Density}
The instanton density for SU(N) has been computed in ref.~[\ref{Bernard}].
In the absence of fermions it is given by the formula
\beq
 \label{densityPV}
 W^{\rm PV}
 = \frac{4}{\pi^2}\frac{\exp[-\alpha(1)-2(N-2)\alpha(\half)]}{(N-1)!(N-2)!}
\times   \int \frac{d^4zd\rho}{\rho^5} \left(\frac{4\pi^2}{g^2}\right)^{2N}
\exp\left[-\frac{8\pi^2}{g^2(\rho)} \right],
\eeq
where $g$ is the strong coupling constant,
and (see ref.~[\ref{tHooftPRD}])
\beqn
\alpha(1)&=&0.443307 \\
\alpha\left(\frac{1}{2}\right)&=&0.145873.
\eeqn
The above formula is given
in the Pauli-Villars regularization scheme, with
\beq
\frac{8\pi^2}{g^2(\rho)}=\frac{8\pi^2}{g^2}-\frac{11 N}{3}\log(m_0\rho),
\eeq
where $m_0$ is the Pauli-Villars mass, and $g$ without argument is the
bare coupling.
If we want to use the value of $\alpha_S$ measured
in today's high-energy experiments, we should convert the above formula
to the $\overline{\rm MS}$ scheme. This change of scheme was first given
in ref.~[\ref{tHooftPRD}], and then corrected in ref.~[\ref{tHooftPREP}].
We have also computed the change of scheme by just computing the
vacuum polarization in the background gauge, in both the Pauli-Villars
and the $\overline{\rm MS}$ scheme. We find that the two schemes
give the same result if the Pauli-Villars mass $m_0$ and the $\overline{\rm
MS}$
scale are related by the formula
\beq
\frac{11}{3}\log\frac{m_0^2}{\mu^2}-\frac{1}{3}=0.
\eeq
Expressing $m_0$ as a function of $\mu$ and replacing it in
formula~(\ref{densityPV}) we obtain
\beq
W^{\overline{\rm MS}}
 = \frac{4}{\pi^2}\frac{\exp
\left[\frac{N}{6}-\alpha(1)-2(N-2)\alpha\left(\frac{1}{2}\right)\right]
}{(N-1)!(N-2)!}
    \int \frac{d^4zd\rho}{\rho^5} \left(\frac{4\pi^2}{g^2}\right)^{2N}
\exp\left[ -\frac{8\pi^2}{g^2(\rho)} \right],
\eeq
which agrees with ref.~[\ref{tHooftPREP}] in the SU(2) case.
Including the fermions, we obtain the extra factor
\beq
\rho^{\nf}\Pi_i m_i \exp\left[-\frac{2}{3}\nf
\log(m_0\rho)+2\nf\alpha\left({\scriptstyle \frac{1}{2}} \right) \right].
\eeq
For the fermion contribution to the vacuum polarization, we find
that the Pauli-Villars and $\overline{\rm MS}$ results agree if
$\mu=m$, so that in this case it is enough to replace $m_0$ with $\mu$.
We will also need to express the running mass in terms of invariant
mass parameters
\beq
m(\mu)=\hat{m}\left(\log\frac{\mu}{\Lambda}\right)^{-\frac{12}{33-2\nf}}
\eeq
according to the definition of ref.~[\ref{Braaten}].
Observe that consistency of the order at which we are computing requires
that one uses the full two-loop expression for $g$ in the exponent
(this was not included in ref.~[\ref{Ellis}])
\beq
\frac{2\pi}{\alpha_S(\rho)}=4\pi b_0
\log\left(\frac{1}{\Lambda \rho}\right)
\left[
    1+\frac{6(153-19 \nf)}{(33-2\nf)^2}
\frac{\log\log\frac{1}{\Lambda^2\rho^2}}{ \log\frac{1}{\Lambda^2\rho^2}}
\right],
\eeq
while in the prefactor a leading-order formula is accurate enough.
Gathering all the factors, and setting $\mu=1/\rho$, we get
\beqn
D(\rho) &=& H \left[\log\frac{1}{\rho^2\Lambda^2}\right]^c
\rho^{6+\frac{\nf}{3}} \\
H &=&
\left(\Pi_i \frac{\hat{m}_i}{\Lambda}\right) \Lambda^{11+\frac{\nf}{3}}
\frac{2}{\pi^2} \exp\left[-\alpha(1)+\half
+(2\nf-2)\alpha\left(\half\right) \right] \nonumber \\
&\times& \left(\frac{33-2\nf}{12}\right)^6  2^{\frac{12 \nf}{33-2\nf}}
  \\
c&=&\frac{45-5 \nf}{33-2\nf}.
\eeqn
Our final formula for the instanton contribution to the vacuum polarization
is then
\beqn
&& \Pi_{\mu\nu}(p^2)=\Pi^0_{\mu\nu}(p^2)
+ K_0
 \left(\Pi_i \frac{\hat{m}_i}{\Lambda}\right)
\left(\frac{p^2}{\Lambda^2}\right)^{-\frac{33+\nf}{6} }
\left[\log\frac{p^2}{\Lambda^2}\right]^\frac{45-5 \nf}{33-2\nf}
\phantom{aaaaaaaaaaaaaaaa}\nonumber \\
&&\times \left\{
\left[C\left(10+\frac{\nf}{3}\right) -1 \right]
(p^2 g_{\mu\nu}-p_\mu p_\nu)
+(C-1)\left(10+\frac{\nf}{3}\right) p_\mu p_\nu
\right\}
\eeqn
 with
\beqn
K_0 &=& \frac{2}{\pi^2} \exp\left[-\alpha(1)+\half
+(2\nf-2)\alpha\left(\half\right) \right]
\left(\frac{33-2\nf}{12}\right)^6
  2^{\frac{12 \nf}{33-2\nf}} \nonumber \\
&\times& \frac{\left(9+\frac{\nf}{3}\right)
\Gamma\left(\frac{3}{2}\right) \Gamma^3\left(\frac{9}{2}+\frac{\nf}{6}\right)}
{\left(7+\frac{\nf}{3}\right)\Gamma\left(6+\frac{\nf}{6}\right)} \\
 C&=&\pm (m_u/m_d+m_d/m_u)/2
\eeqn
where in the last expression the $+$ sign is appropriate for vector-vector,
and the $-$ sign for axial-axial correlators.
The Born term (which fixes our normalization) is given by
\beq
\Pi_{\mu\nu}^0=\frac{1}{4\pi^2}(p_\mu p_\nu - p^2 g_{\mu\nu})\log p^2
\eeq
both for the axial and the vector contribution.

All results quoted so far were obtained in the Euclidean
metric. The corresponding Minkowski space formulae in the time-like
region are obtained by analytic continuation
in $p^2$. Observe that in the case of the electromagnetic current,
which has $C=1$ the sign of the instanton correction is opposite
to that of the leading term, contrary to the result of
ref.~[\ref{Ellis}].

\section{Final Results}
It is now straightforward to obtain the instanton contribution to
the $\tau$ hadronic decays. According to
ref.~[\ref{Braaten}], using the same notation,
the ratio of the hadronic to the leptonic width $R_\tau$ is given by
\beq
R_\tau=
6\pi i \int_{\abs{z}=1}dz(1-z)^2
 \left[(1+2z)\Pi^{(T)}_{A+V}(p^2)+\Pi^{(L)}_{A+V}(p^2)\right],
\eeq
where $z=p^2/M_\tau^2$, and
\beq
\Pi^{\mu\nu}_{A+V}=\Pi^{(T)}_{A+V}(p^2)(p^\mu p^\nu - p^2 g^{\mu\nu})
+\Pi^{(L)}_{A+V}(p^2) p^\mu p^\nu.
\eeq
The suffix $A+V$ indicates the sum of the axial and vector
contributions.
We get
\beq
\frac{R^{\;	\rm inst}_\tau}{R^{0}_\tau}=
-2\pi i K_0 \left(\frac{\Lambda}{M_\tau}\right)^9
\frac{\hat{m}_u \hat{m}_d \hat{m}_s}{M_\tau^3}
\int_{\abs{z}=1} dz(1-z)^2 z^{-6}
\left(\log\frac{-z M_\tau^2}{\Lambda^2}\right)^\frac{10}{9}
\left[2(1+2z)-22\right].
\eeq
Observe now that the contour integral would give zero if the logarithmic
term was not present. Since the power of the logarithm is very near 1,
we expect that the integral will not depend much upon the ratio
$\Lambda/M_\tau$. In fact
we find the numerical result
\beq
\frac{R^{\rm inst}_\tau}{R^{0}_\tau}
=\left(\frac{3.64\; \Lambda}{M_\tau}\right)^9
\frac{\hat{m}_u\hat{m}_d\hat{m}_s}{M_\tau^3},
\eeq
where the coefficient 3.64 corresponds to $\Lambda=400\,$MeV; it
varies from 3.67 to 3.62 if $\Lambda$ is pushed to the
extreme values of 0.1 and 1~GeV.
Choosing with ref.~[\ref{Braaten}]
\beq
\hat{m}_u=8.7\,\mbox{\rm MeV},\quad \hat{m}_d=15\,\mbox{\rm MeV}\;
\;\mbox{\rm and}\;\;
\hat{m}_s=270\,\mbox{\rm MeV}
\eeq
with $M_\tau=1.784$~GeV we get the result
\beq
\frac{R^{\rm inst}_\tau}{R^{0}_\tau}
=\left(\frac{0.96 \times \Lambda}{1.784\; \mbox{GeV}}\right)^9
\eeq

The value of $\Lambda$ to be used in this context is $\Lambda_3$, which
is somewhat larger than the corresponding values of $\Lambda_5$ that
are usually quoted. For example, a recent review~\refq{\ref{Altarelli}}
quotes the range $150<\Lambda_5<330\,$MeV, which corresponds
roughly to $280<\Lambda_3<510\,$MeV. We see that even in the most pessimistic
case of the largest allowed value for $\Lambda_3=510\,$MeV the instanton
correction would turn out to be absolutely negligible.
\section{Conclusions}
We have completed a calculation of the one-instanton contribution
to the $\tau$ hadronic width. We found that the corrections
are actually negligible.
Part of the smallness of the result is due to the fact that
the one-instanton correction is proportional to the chiral suppression
factor $m_u m_d m_s/M_\tau^3$. It is interesting to note that
if that factor were not there, the instanton correction could be
of order 1, and would thereby invalidate the conclusions of
ref.~[\ref{Braaten}].
At this point one may wonder whether corrections due to
instanton anti-instanton pairs, which should be suppressed by
18 powers of the ratio $\Lambda/M_\tau$, but do not carry any chiral
suppression, may have coefficients of comparable
size, that is to say, if they become of order 1 for
$M_\tau\approx 4\Lambda_3$. The computation of ref.~[\ref{Balitsky}]
seems to indicate large corrections, although one may doubt
the reliability of the method used there to perform a computation
beyond the dilute gas approximation.

Our result for the instanton correction to the vacuum polarization,
for $n_{\rm f}=3$, can
be summarized as follows
\beqn
\Pi_{\mu\nu}(p^2)&=&
\frac{1}{4\pi^2}\Bigg\{ (p_\mu p_\nu - p^2 g_{\mu\nu})\log p^2
+ \frac{\hat{m}_u \hat{m_d} \hat{m_s}}{p^3}
\left(\frac{5.1701\; \Lambda}{p}\right)^9
\left[\log\frac{p^2}{\Lambda^2}\right]^\frac{10}{9} \nonumber \\
&&\times \left[
\left( {\scriptstyle \frac{11}{10}} C - {\scriptstyle \frac{1}{10}} \right)
(p^2 g_{\mu\nu}-p_\mu p_\nu)
+ {\scriptstyle \frac{11}{10}}(C-1) p_\mu p_\nu
\right] \Bigg\}
\eeqn
with $C=\pm(m_u/m_d+m_d/m_u)$. For electromagnetic currents ($C=1$),
approximating the logarithmic exponent with 1, we get
\beq
\frac{R^{\rm inst}_\gamma}{R^{(0)}_\gamma}=
\left(\frac{Q_0}{p}\right)^{12}
\eeq
with $Q_0=0.591\;$GeV for $\Lambda_3=300\;$MeV and
$Q_0=0.867\;$GeV for $\Lambda_3=500\;$MeV.
The considerable numerical differences with respect to ref.~[\ref{Ellis}]
have several origins.
First of all, there was an error in the
normalization factor for the instanton density\refq{\ref{tHooftPRD}},
which was subsequently corrected by the author himself.
The authors of ref.~[\ref{Ellis}] use a leading-order expression for $\as$
(instead of a next-to-leading one)
in the exponent of the instanton density, which leads to an
overestimate of the effect. The other differences
are due to the use of the $\MSB$ instead of
the Pauli-Villars scheme in the instanton density,
the different definition
of the invariant quark mass used in the present work, and
the different colour factor we found. These last three differences
have a minor impact on the final result.
\eject
\vskip 1cm
{\bf REFERENCES}
\begin{enumerate}
\item\label{Braaten}
   E. Braaten, S. Narison and A. Pich, \np{B373}{92}{581},\newline
   F. Le Diberder and A. Pich, \pl{B289}{92}{165}.
\item\label{Shifman}
 M.A. Shifman, A.L. Vainshtein and V.I. Zakharov, \np{B147}{79}{385}, 448, 519.
\item\label{Mueller}
   A.H. Mueller, \np{B250}{85}{327}.
\item\label{Sakarov}
   V.I. Zakharov, \np{B385}{92}{452}; \newline
   M. Beneke and V.I. Zakharov, Munich preprint MPI-Ph/92-53 (1992).
\item\label{Gross}
   N. Andrei and D.J.Gross, \pr{D18}{78}{468}.
\item\label{Ellis}
   L. Baulieu \etal, \pl{B77}{78}{290}.
\item\label{Bernard}
   C. Bernard, \pr{D19}{79}{3013}.
\item\label{tHooftPRD}
   G. 't Hooft, \pr{D14}{76}{3432}.
\item\label{tHooftPREP}
   G. 't Hooft, \prep{142}{86}{357}.
\item \label{Altarelli}
   G. Altarelli, CERN-TH.6623/92, talk given at the Workshop
  ``QCD: 20 Years Later'', Aachen, Germany, 9--13 June 1992.
\item \label{Balitsky}
   I.I. Balitsky, \pl{B273}{91}{282}.
\end{enumerate}
\end{document}